\documentclass[acmlarge]{acmart}

\AtBeginDocument{%
  }

\setcopyright{acmcopyright}
\copyrightyear{2018}
\acmYear{2018}
\acmDOI{XXXXXXX.XXXXXXX}

\acmJournal{POMACS}
\acmVolume{37}
\acmNumber{4}
\acmArticle{111}
\acmMonth{8}

\usepackage{amsfonts}
\usepackage{amsmath}
\usepackage{bbm}
\usepackage{graphicx}
\usepackage{xspace}
\usepackage{mathtools,soul}
\usepackage[flushleft]{threeparttable}
\usepackage{multirow}
\usepackage{rotating}
\usepackage{booktabs}	
\usepackage{wrapfig, lipsum}
\usepackage{subcaption}
\usepackage{enumitem}

\graphicspath{ {./plots/entropy/} {./plots/sim/} {./plots/stability/} {./plots/runtimeLayers/} {./figures/} {./plots/stabilityDims/}}
\newcommand{\etal}{\textit{et al}. }

\newcommand{\bo}[1]{{\small\color{blue}{\xspace#1}}}
\newcommand{\xia}[1]{{\small\color{red}{\xspace#1}}}

\newcommand{\method}{\mbox{$\mathop{\mathtt{RAM}}\limits$}\xspace}
\newcommand{\methodmu}{\mbox{$\mathop{\mathtt{RAM\text{-}u}}\limits$}\xspace}
\newcommand{\SASRec}{\mbox{$\mathop{\mathtt{SASRec}}\limits$}\xspace}
\newcommand{\Caser}{\mbox{$\mathop{\mathtt{Caser}}\limits$}\xspace}
\newcommand{\HGN}{\mbox{$\mathop{\mathtt{HGN}}\limits$}\xspace}
\newcommand{\HAM}{\mbox{$\mathop{\mathtt{HAM}}\limits$}\xspace}
\newcommand{\NARM}{\mbox{$\mathop{\mathtt{NARM}}\limits$}\xspace}
\newcommand{\GRURec}{\mbox{$\mathop{\mathtt{GRU4REC}}\limits$}\xspace}
\newcommand{\GRURecP}{\mbox{$\mathop{\mathtt{GRU4REC\text{++}}}\limits$}\xspace}
\newcommand{\NextItRec}{\mbox{$\mathop{\mathtt{NextItRec}}\limits$}\xspace}

\newcommand{\Beauty}{\mbox{$\mathop{\texttt{Beauty}}\limits$}\xspace}
\newcommand{\Toys}{\mbox{$\mathop{\texttt{Toys}}\limits$}\xspace}
\newcommand{\Children}{\mbox{$\mathop{\texttt{Children}}\limits$}\xspace}
\newcommand{\Comics}{\mbox{$\mathop{\texttt{Comics}}\limits$}\xspace}
\newcommand{\MLOM}{\mbox{$\mathop{\texttt{ML-1M}}\limits$}\xspace}
\newcommand{\MLTM}{\mbox{$\mathop{\texttt{ML-20M}}\limits$}\xspace}

\begin{document}

\title{Recursive Attentive Methods with Reused Item Representations 
	for Sequential Recommendation}

\author{Bo Peng}
\email{peng.707@buckeyemail.osu.edu}
\affiliation{%
  \institution{The Ohio State University}
  \state{Ohio}
  \country{USA}
}

\author{Srinivasan Parthasarathy}
\email{srini@cse.ohio-state.edu}
\affiliation{%
  \institution{The Ohio State University}
  \city{Columbus}
  \state{Ohio}
  \country{USA}
}

\author{Xia Ning}
\authornote{Corresponding author}
\email{ning.104@osu.edu}
\affiliation{%
  \institution{The Ohio State University}
  \city{Columbus}
  \state{Ohio}
  \country{USA}
}

\renewcommand{\shortauthors}{Peng et al.}

\begin{abstract}
  Sequential recommendation aims to recommend the next item of users' interest based on their historical interactions.
  Recently, the self-attention mechanism has been adapted for sequential recommendation, 
  and demonstrated state-of-the-art performance.
  However, in this manuscript, we show that the self-attention-based sequential recommendation methods could suffer from the localization-deficit issue.
  As a consequence, 
  in these methods, 
  over the first few blocks, 
  the item representations may quickly diverge from their original representations, 
  and thus, impairs the learning in the following blocks. 
  %
  To mitigate this issue, 
  in this manuscript, we develop a 
  recursive attentive method with reused item representations (\method) 
  for sequential recommendation.
  We compare \method with five state-of-the-art
  baseline methods on six public benchmark datasets. 
  Our experimental results demonstrate that \method significantly outperforms the baseline methods on benchmark datasets, with an improvement of as much as 11.3\%. 
  Our stability analysis shows that \method could enable deeper and wider models for better performance.
  Our run-time performance comparison signifies that \method could also be more efficient on benchmark datasets.
\end{abstract}



\maketitle

\section{Introduction}
\label{sec:introduction}

Sequential recommendation aims to 
identify and recommend the next item of users' interest based 
on their historical interactions.
It has been widely employed in applications such as online shopping~\cite{he2016ups,mcauley2015image} 
and video streaming~\cite{belletti2019quantifying}, 
and has been drawing increasing attention from the research community.
%
%
Recently, Self-Attention (SA) based methods~\cite{sasrec} have been developed for sequential recommendation, 
and demonstrated state-of-the-art performance. 
SA-based methods usually stack 
multiple SA blocks to learn users' preferences. 
%
In each block, for an item in the history, 
these methods use the SA 
mechanism to aggregate 
semantically relevant items, 
and update the representation of 
the given item to a better one.
%
To recursively improve the quality
of the item representations through blocks,
a localized attention map~\cite{shim2021understanding}  
focusing on a few semantically relevant items
is required in each block.
However, 
as will be shown in Section~\ref{sec:motivation}, 
in sequential recommendation, 
the learned attention maps from SA-based methods could suffer 
from the localization-deficit issue.
As a consequence, 
in the first few blocks, 
the item representations could quickly diverge from their original ones, 
and thus, further impairs the learning of the following blocks and item embeddings.
Through multiple blocks, this problem could be amplified, 
and eventually deteriorate the recommendation performance.
%

To avoid the above issue,
in this manuscript, we reuse the item representations in all the blocks 
(i.e., without update), and develop a novel 
\ul{R}ecursive \ul{A}ttentive \ul{M}ethod with reused item representations, 
denoted as \method,  
for sequential recommendation.  
In \method, we stack attention blocks to recursively 
capture users' short-term preferences 
with same, fixed item representations among all the blocks, 
and also explicitly learn user embeddings to capture users' long-term preferences.
%
%
%
We compare \method with five state-of-the-art baseline methods 
on six benchmark datasets.
Our experimental results demonstrate that 
on the six datasets, overall, 
\method could significant outperform 
the state-of-the-art baseline methods, 
with an improvement of up to 11.3\%.
Notably, over the SA-based baseline method, 
\method and its variant could achieve a significant improvement of up 
to 7.1\% over the six datasets. 
%
%
%
%
In addition, our stability analysis shows that 
\method could be more stable than SA-based methods on 
learning deep and wide models for better performance.
Our run-time performance comparison demonstrates that 
\method could also be more efficient than 
SA-based methods on all the benchmark datasets.

We summarize our major contributions as follows:
\begin{itemize}[noitemsep,topsep=0pt]
	\item We identify the localization-deficit issue in SA-based sequential recommendation methods. 
	To mitigate the effects of this issue, 
	we reuse item representations through all the blocks 
	and develop the novel \method method for sequential recommendation.
	\item \method significantly outperforms five state-of-the-art baseline 
	methods on six benchmark datasets.
	\item Our user embedding analysis shows the importance of 
	explicitly learning users' long-term preferences for sequential recommendation.
	\item Our stability analysis reveals that \method is more stable than SA-based sequential recommendation methods for learning deep and wide models.
	\item Our run-time performance comparison demonstrates that \method is also more efficient than SA-based methods. 
\end{itemize}

\section{Related Work}
\label{sec:literature}

\subsection{Sequential Recommendation}
\label{sec:literature:sequential}

%
In the last few years, neural networks 
such as Recurrent Neural Networks (RNNs) and Convolutional Neural Networks (CNNs)  
have been widely adapted for sequential recommendation.
For example, Hidasi~\etal~\cite{hidasi2015session} 
developed a Gated Recurrent Units (GRUs) based method \GRURec 
in which GRUs are employed to recurrently model users' preferences.
Hidasi~\etal~\cite{hidasi2018recurrent} developed \GRURecP, which
uses a novel ranking loss to mitigate the 
degradation problem in GRUs when processing long sequences.
Vasile~\etal~\cite{vasile2016meta} developed a skip-gram-based method 
to leverage the skip-gram model~\cite{mikolov2013distributed} to capture the co-occurrence among items for recommendation.
Recently, CNN's are also adapted 
for sequential recommendation.
Tang~\etal~\cite{tang2018personalized} developed a CNNs-based model \Caser, 
which employs vertical and horizontal convolutional filters to capture 
the synergies among items for a better recommendation.
Yuan~\etal~\cite{yuan2018simple} developed a CNNs-based generative model \NextItRec 
to model the long-term preferences of users in CNNs.
Besides neural networks, attention mechanisms are also widely used 
in developing sequential recommendation methods.
Li~\etal~\cite{li2017neural} developed \NARM, which 
incorporates attention mechanisms into RNNs 
to better capture users' long-term preferences.
Kang~\etal~\cite{sasrec} developed a SA-based method \SASRec, 
which stacks multiple SA blocks 
to recursively learn users' preferences.
Ma~\etal~\cite{ma2019hierarchical} developed \HGN, 
which uses attention mechanisms to identify important items and learn
useful latent features for the items for better recommendations.
%
In addition, 
simple pooling mechanisms are also adapted for sequential recommendation.
Peng~\etal~\cite{peng2021ham} developed hybrid associations models (\HAM) 
in which pooling mechanisms are employed to learn the 
associations and synergies 
among items.
 
\subsection{Self Attention}
\label{sec:literature:attention}

The self-attention (SA) mechanism~\cite{vaswani2017attention} is originally developed for machine translation~\cite{bahdanau2014neural}, 
and later has also been demonstrated to achieve the 
state-of-the-art performance on other Nature Language Processing (NLP) tasks such as speech recognition~\cite{shim2021understanding,zhang2021usefulness} and text generation~\cite{tay2021synthesizer}. 
With the prosperity of SA-based methods, 
recently, a lot of research has been done to 
investigate the SA mechanisms and the learned attention weights. 
Tay~\etal~\cite{tay2021synthesizer} 
investigated the dot-product attention in SA, 
and found that it 
may not be necessary for machine translation and text generation, among other tasks.
Zhang~\etal~\cite{zhang2021usefulness} investigated 
the attention weights learned in different SA blocks, 
and found that for speech recognition, 
in the upper blocks, 
the weights focus on the diagonal 
of the attention map, and thus, 
each frame primarily aggregates 
the information from near frames for recognition.
Shim~\etal~\cite{shim2021understanding} also investigated the 
attention weights learned for speech recognition, 
and found that SA could learn phonetic features in the lower blocks, 
and linguistic features in the higher blocks.
Although widely studied in NLP tasks, 
to the best of our knowledge, 
in sequential recommendation, 
the attention weights learned in SA are still unexplored.
In this manuscript, we analyze the attention weights in SA and identify potential issues (Section~\ref{sec:motivation}). 

\section{Definition and notations}
\label{sec:notation}

\begin{table}[!h]
  \vspace{-15pt}
  \caption{Key Notations}
  \label{tbl:notations}
  \centering
  \begin{threeparttable}
     \begin{footnotesize}
      \begin{tabular}{
	@{\hspace{3pt}}l@{\hspace{3pt}}
	@{\hspace{3pt}}p{0.35\textwidth}@{\hspace{3pt}}          
	}
        \toprule
        notations & meanings \\
        \midrule
        $\mathbb{U}$    &  the set of users \\
        $\mathbb{V}$    &  the set of items \\
        $S_i$		&  the historical interaction sequence of user $u_i$ \\
        $B_i$    & the fixed-length sequence converted from $S_i$\\
        $n$   & the number of items in $B_i$\\
        $d$     &  the dimension of  embeddings\\
        $\textbf{h}^{(m)}_i$ & the representation of the short-term preferences of user $u_i$ from the $m$-th attention block\\
        $\textbf{u}_i$ & the embedding of user $u_i$\\        
        $\hat{r}_{ij}$ & the recommendation score of user $u_i$ on item $v_j$\\
        \bottomrule
      \end{tabular}
      \end{footnotesize}
  \end{threeparttable}
  \vspace{-10pt}
\end{table}

In this manuscript, we tackle the sequential recommendation problem that, given the historical interactions of users, we recommend the next item of users' interest.
In this manuscript, $\mathbb{U} = \{u_1, u_2, \dots\}$ is the set of all the users, where $u_i$ represents a certain user, and $|\mathbb{U}|$ is the total number of users in the dataset.
Similarly, we represent the set of all the items as $\mathbb{V} = \{v_1, v_2, \dots\}$, 
and $|\mathbb{V}|$ is the total number of items in the dataset.
The historical interactions of a user $u_i$ is represented as a sequence $S_i = \{s_1(i), s_2(i), \dots\}$, where $s_t(i)$ is the $t$-th interacted item in the history and $|S_i|$ is the length of the sequence.
When no ambiguity arises, we will eliminate $i$ in $S_i$. 
We use uppercase letters to denote matrices, lower-case and bold letters to
denote row vectors and lower-case non-bold letters to represent scalars. 
Table~\ref{tbl:notations} presents the key notations used in the manuscript.

\section{Localization-Deficit Issue in Self-Attention for Sequential Recommendation} 
\label{sec:motivation}
%
As defined in the literature~\cite{shim2021understanding},
localization denotes the role of attention maps 
focusing on
semantically relevant elements to extract meaningful features.
A recent study in speech recognition~\cite{shim2021understanding} 
shows that the localization of attention maps enables SA to effectively 
aggregate semantically relevant frames and learn clustered phonetic and linguistic features through the SA blocks.
%
%
Although widely studied in speech recognition ~\cite{shim2021understanding,zhang2021usefulness,yang2020understanding}, in sequential recommendation, to the best of our knowledge, the localization of attention maps in SA is still unexplored.
%
%
%
To fill this research gap, in this manuscript, 
we analyze the localization of attention maps in SA for sequential recommendation.
%
%

%
Specifically, 
we export the attention maps learned from the state-of-the-art SA-based method \SASRec~\cite{sasrec}
on six benchmark 
datasets: \Beauty, \Toys, \Children, \Comics, \MLOM and \MLTM (Section~\ref{sec:materials:datasets}).
%
%
%
In \SASRec, the interaction sequence $S_i$ is converted to a fixed-length sequence $B_i = \{b_1(i), \dots b_n(i)\}$,
 which contains the last $n$ items in $S_i$. 
When no ambiguity arises, we will eliminate $i$ in $B_i$ and $b_t(i)$.
If $|S_i|$ is shorter than $n$, we pad empty items, denoted as $b_0$, 
at the beginning of $B_i$.
%
%
In the $m$-th SA block, 
\SASRec has an attention map $A^{(m)} \in \mathbb{R}^{|\mathbb{U}| \times n \times n}$, 
in which $A^{(m)}_{ijk}$ 
is the attention weight on $b_k(i)$ for $b_j(i)$
(i.e., the attention weight on $b_k(i)$ that is used to update the representation of $b_j(i)$). 
To enable causality~\cite{sasrec}, each item could only attend to itself and all the previous items 
in $B_i$, 
and thus, $A^{(m)}_{ijk}=0$ if $j < k$. 
%

Intuitively, to enable localization, the attention map should satisfy two properties: 
1) the attention weights should be concentrated (i.e., focus on a few items); and 
2) if $A^{(m)}_{ijk}$ is higher, item $b_k(i)$ should be more semantically relevant to $b_j(i)$.
%
In sequential recommendation, the ground-truth relevance among items is usually intractable.
Thus, in this manuscript, we focus on investigating only if the attention weights are concentrated. 
%
In $A^{(m)}$, the attention weights from each item in a user sequence 
are normalized via the softmax function and form a distribution.
Therefore, as suggested in the literature~\cite{entropy}, we could use entropy to measure the concentration level of the attention weights quantitatively. 
%
A lower entropy indicates that the attention weights concentrate on fewer items and thus, a higher concentration level.
Given 
$A^{(m)}$,  
we calculate the average entropy of the attention weights 
for all the items 
as follows: 
\begin{equation}
	\label{eqn:entropy}
	\frac{-1}{\sum \nolimits_{u_i \in \mathbb{U}} \sum \nolimits_{j=1}^n \mathbbm{1} (b_j(i) \ne b_0)} \sum \nolimits_{u_i \in \mathbb{U}} \sum \nolimits_{j=1}^{n} \mathbbm{1} (b_j(i) \ne b_0) \sum \nolimits_{k=1}^{j} (A^{(m)}_{ijk} \log(A^{(m)}_{ijk})), 
\end{equation}
where $\mathbbm{1}(x)$ is an indicator function (i.e., $\mathbbm{1}(x) = 1$ if $x$ is
true, otherwise 0),  
$b_j(i) \ne b_0$ excludes the padding items,
and $-\sum \nolimits_{k=1}^{j} (A^{(m)}_{ijk} \log(A^{(m)}_{ijk}))$ is the entropy of the attention weights from $b_j(i)$.
%
%

%
We compare the average entropy in $A^{(m)}$ with that of weights randomly sampled from a uniform distribution between 0 and 1.
%
%
%
Particularly, for each non-zero entry in $A^{(m)}$, 
we randomly sample a weight from the uniform distribution and normalize the sampled weights 
using the softmax function in the same way as in SA.
We denote the matrix with the random weights as $R$, 
and calculate the average entropy in $R$ using Equation~\ref{eqn:entropy}.
Statistically, the uniformly sampled weights should not focus on any items, 
and thus, have a low level of concentration and a high entropy.

%
%

\begin{figure}
	\centering
	\vspace{-15pt}
	\scalebox{0.8}{
	\input{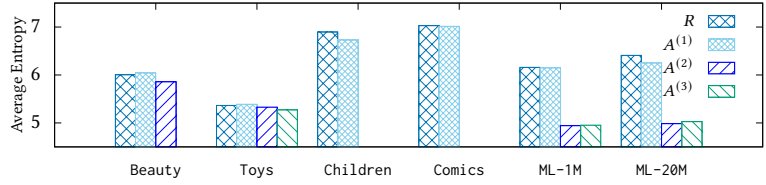}
	}
	\vspace{-15pt}
	\caption{The average entropy in $A^{(1)}$, $A^{(2)}$, $A^{(3)}$ and $R$ from \SASRec on the six datasets.}
	\label{fig:entropy}
	\vspace{-20pt}
\end{figure}

Figure~\ref{fig:entropy} 
shows the average entropy in $R$, $A^{(1)}$, $A^{(2)}$ and $A^{(3)}$  on the six datasets.
%
On these datasets, 
the best performing \SASRec models have up to three SA blocks.
Therefore, for each dataset, we present the average entropy 
in $A^{(1)}$, $A^{(2)}$ and $A^{(3)}$, if available.  
%
%
%
As shown in Figure~\ref{fig:entropy}, 
on all the six datasets, in the first attention block, 
the average entropy in $A^{(1)}$ is very close to that in $R$ (i.e., difference $<3\%$).
%
On \Beauty and \Toys, the average entropy in $A^{(2)}$ and $A^{(3)}$, if available, 
is still close to that in $R$.
These results reveal that for sequential recommendation, 
in SA blocks, especially in the first blocks, 
the learned attention maps may not have 
concentrated weights, 
and thus, cannot enable localization.
We denote this as the localization-deficit issue in SA.
Ideally, we expect the attention weights to converge to a few items after several SA blocks. 
However, as will be demonstrated later in Section~\ref{sec:results:stability:layer}, 
SA-based sequential recommendation methods cannot afford 
very deep architectures; 
item representations aggregated using initially unconcentrated 
attention weights can quickly diverge
from their original representations, 
resulting in more difficulties for the 
following attention weights to concentrate. 
Through multiple blocks, this issue could be amplified, and thus, 
eventually, deteriorate the recommendation performance.

We notice that 
on \MLOM and \MLTM, the average entropy in $A^{(2)}$ 
and $A^{(3)}$ is considerably lower than that in $R$, 
which may indicate that the attention weights 
in these attention maps could be concentrated.
%
As will be shown in Section~\ref{sec:materials:datasets}, 
\MLOM and \MLTM are the two most dense datasets among the six datasets, 
which could afford learning of concentrated attention weights in the last few blocks.
%
%
However, even on the densest \MLOM and \MLTM datasets, 
in the first block, 
the learned attention weights 
still have low concentration, 
indicating the general ubiquity of localization-deficit issues in sequence recommendation.
In this manuscript, 
instead of investigating the fundamental mechanism 
leading to localization deficit in sequential recommendation, 
we focus on mitigating the effects of this issue and 
improve sequential recommendation performance while still 
using attention blocks (not SA blocks). 
Our strategy is to reuse item representations, 
that is, fix item representations throughout all the attention blocks and 
not update them using attention weights. 
We develop the novel recursive attentive method with reused item representations, denoted as \method, to implement
this strategy for sequential recommendation. 
As will be shown in Section~\ref{sec:results}, 
compared to existing SA-based methods, 
\method can achieve better recommendation performance, lower time complexity, and more stability in learning deep 
and wide models.

\section{Method}
\label{sec:method}


\begin{figure}[t]
	\includegraphics[width=0.85\linewidth, angle=270, scale=0.3]{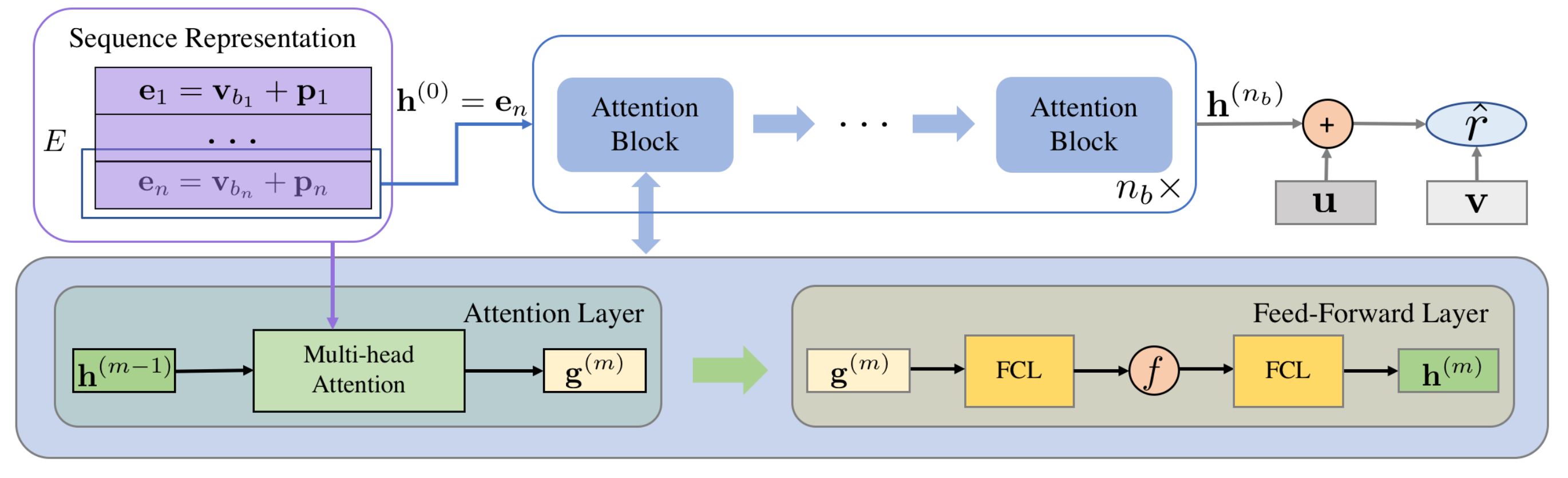}
	\vspace{-10pt}
	\caption{Overall Architecture for \method}
	\label{fig:architecture}
	\vspace{-15pt}
\end{figure}

Figure~\ref{fig:architecture} presents the overall architecture of \method. 
In \method, we stack multiple attention blocks with reused item representations to model users' short-term preferences, 
and explicitly learn user embeddings to capture long-term preferences.
%
%
We present the detail of \method in the following sections.

\subsection{Sequence Representation}
\label{sec:method:seq}

Following \SASRec, we focus on the most recent $n$ items in users' interaction histories to generate the recommendations.
Specifically, we transform each interaction sequence $S$ to a fixed-length sequence $B$
as presented in Section~\ref{sec:motivation}.
%
%
%
In \method, following the literature~\cite{sasrec,vaswani2017attention}, we represent items and positions in each sequence using learnable embeddings.
We learn an item embedding matrix $V \in \mathbb{R}^{|\mathbb{V}| \times d}$ in which the $j$-th row $\mathbf{v}_j$ 
is the embedding of item $v_j$,  
and $d$ is the dimension of embeddings.
We use a constant zero vector as the embedding of padding items.
Similarly, we learn a position embedding matrix $P \in \mathbb{R}^{n \times d}$, and $\mathbf{p}_t$ is the embedding of the $t$-th position.
Given $V$ and $P$, we represent the sequence $B$ using a matrix $E$ as follows:
\begin{equation}
	\label{eqn:seq}
	E = [\mathbf{e}_1; \mathbf{e}_2; \cdots; \mathbf{e}_n] =  [\mathbf{v}_{b_1}+\mathbf{p}_{1}; \mathbf{v}_{b_2}+\mathbf{p}_{2}; \cdots; \mathbf{v}_{b_n}+\mathbf{p}_{n}],
\end{equation}
where $\mathbf{e}_t = \mathbf{v}_{b_t}$+$\mathbf{p}_{t}$ is the $t$-th row in $E$, 
and $\mathbf{v}_{b_t}$ is the embedding of the $t$-th item in $B$.

\subsection{Modeling Users' Short-term Preferences}
\label{sec:method:attention}

SA-based methods stack multiple SA blocks to recursively 
model users' preferences, and have been demonstrated~\cite{peng2021ham,sasrec} the state-of-the-art performance in sequential recommendation.
Following the idea, in \method, we also stack multiple attention blocks to 
model users' preferences in a recursive manner.
%
%
In addition, we reuse the item representations 
in all the attention blocks to mitigate localization deficit. 
%
%
In \method, we have $n_b$ attention blocks, 
and each attention block contains an attention layer  
and a feed-forward layer.

\subsubsection{Attention layer with reused item embeddings}
\label{sec:method:attention:attention}

In the attention layer, we employ the multi-head attention mechanism~\cite{vaswani2017attention} to refine 
the hidden representation of users' preferences from the previous block as follows:
\begin{eqnarray}
	\label{eqn:multi}
	\begin{aligned}
		\boldsymbol{\beta}^{(m)}_k &= \text{softmax}(\frac{(\mathbf{h}^{(m-1)} Q^{(m)}_k) (E Z^{(m)}_k)^\top}{\sqrt{d}}), \\
		\mathbf{head}^{(m)}_k &= \boldsymbol{\beta}^{(m)}_k (E W^{(m)}_k), \\
		\mathbf{g}^{(m)} &= [\mathbf{head}^{(m)}_1, \mathbf{head}^{(m)}_2, \dots, \mathbf{head}^{(m)}_{n_h}] C^{(m)},\\
	\end{aligned}
\end{eqnarray}
where $\mathbf{h}^{(m-1)}$ is the hidden representation of users from the $(m\text{-}1)$-th block, and 
$\mathbf{g}^{(m)}$ is the output of the attention layer in the $m$-th block.
$Q^{(m)}_k, Z^{(m)}_k \text{and } W^{(m)}_k \in \mathbb{R}^{d \times \frac{d}{n_h}}$ are learnable parameters 
in the $k$-th head of the attention layer in the $m$-th block, 
and $C^{(m)} \in \mathbb{R}^{d \times d}$ is also a 
learnable parameter in the $m$-th block.
$\mathbf{head}^{(m)}_k$ is the output of the $k$-th head, and
$n_h$ is the number of heads. 
%
%
%
Note that as in Equation~\ref{eqn:multi}, 
different from the SA layer in SA-based methods, 
in \method, 
we reuse $E$ in all the attention layers, 
%
which could alleviate the effects of cumulating unconcentrated attention weights through blocks. 
%
%
As will be shown in Section~\ref{sec:method:complexity}, 
by doing this, 
we can also significantly improve the time complexity of computing each attention block from $\mathcal{O} (n^2d+nd^2)$ as in SA 
to $\mathcal{O} (nd^2)$, 
where $d$ is the embedding dimension and $n$ is the length of the fixed-length sequences (i.e., $B$).
%
%
%
%

\subsubsection{Feed-forward layer}
\label{sec:method:attention:feed} 

Following the literature~\cite{vaswani2017attention,sasrec,fan2022sequential}, we include a feed-forward layer after the attention layer 
to endow \method with non-linearity, 
and better model users' preferences.
Particularly, given $\mathbf{g}^{(m)}$ as in Equation~\ref{eqn:multi}, 
we stack two fully-connected layers (FCLs) as the feed-forward layer as follows:
\begin{equation}
	\label{eqn:feed}
	\mathbf{h}^{(m)} = (f(\mathbf{g}^{(m)}W^{(m)}_1+\mathbf{b}^{(m)}_1))W^{(m)}_2 + \mathbf{b}^{(m)}_2,
\end{equation}
where $f(\cdot)$ is an activation function such as ReLU~\cite{nair2010rectified} and GELU~\cite{hendrycks2016gaussian}, 
$W^{(m)}_1$, $W^{(m)}_2 \in \mathbb{R}^{d \times d}$ and $\mathbf{b}^{(m)}_1$, $\mathbf{b}^{(m)}_2$ are learnable parameters.
%
Following the literature~\cite{sasrec,vaswani2017attention}, 
we stack attention layers and feed-forward layers using the residual connections~\cite{he2016deep} 
to mitigate the degradation problem during the training.

Similarly to that in \SASRec, we use $\mathbf{e}_n$ 
(i.e., the representation of the last item in $E$) 
as the latent representation of users in the first attention block (i.e., $\mathbf{h}^{(0)} = \mathbf{e}_n$).
Intuitively, $\mathbf{e}_n$ consists of embedding the most recently interacted item, and thus, could be viewed as a representation of users' short-term preferences~\cite{stamp,wu2019session}.
Conditioned on $\mathbf{e}_n$, stacking multiple attention blocks could be viewed as a process 
to refine the learned representation of users' short-term preferences, 
and thus, $\mathbf{h}^{(n_b)}$ (i.e., the output of the last attention block)
could be viewed as the representation of users' short-term preferences.


\subsection{Modeling users' long-term preferences} 
\label{sec:method:long}

%
%
%
%
As demonstrated in the literature~\cite{peng2021ham,peng2020m2}, both users' 
short-term preferences and long-term preferences play important roles in generating accurate recommendations. 
However, as discussed in Section~\ref{sec:method:attention}, 
the attention blocks may focus on learning users' short-term preferences. 
As a result, it may not be able to capture users' long-term preferences fully.
To better capture users' long-term preferences, in \method, following the literature~\cite{peng2021ham,ma2019hierarchical}, 
we explicitly learn user embeddings. 
Specifically, we learn an embedding matrix $U \in \mathbb{R}^{|\mathbb{U}| \times d}$ in which the $i$-th row $\mathbf{u}_i$ is the embedding of user $u_i$.
Note that, as suggested in the literature~\cite{sasrec}, 
by considering all the interacted items in each block,
SA-based methods (e.g., \SASRec) should be able to implicitly 
learn users' long-term preferences, 
and thus, explicitly learning user embeddings may not benefit the performance.
In \method, we also consider all the interacted items in each block. 
However, as will be shown in Section~\ref{sec:results:user}, 
we empirically find that additionally learning user embeddings could 
significantly improve 
the recommendation performance on most of the benchmark datasets.
In addition, as will be shown in Section~\ref{sec:results:user:sim}, 
the learned user embeddings 
and the output from the attention blocks (i.e., $\mathbf{h}^{(n_b)}$, the representation of user's short-term preferences) 
capture different preferences of users, 
and thus, are complementary for better performance.
%
%
%

\subsection{Recommendation Scores and Network Training}
\label{sec:method:scores}

Given the output of the attention blocks and the user embeddings, 
we generate recommendation scores for candidate items as follows:
\begin{equation}
	\label{eqn:score}
	\hat{r}_{ij} = (\mathbf{h}^{(n_b)}_i + \mathbf{u}_i) \mathbf{v}_j^\top, 
\end{equation}
where $\hat{r}_{ij}$ is the recommendation score of user $u_i$ on item $v_j$, 
and $n_b$ is the total number of attention blocks in \method.  
For each user, we recommend items with the top-$k$ highest recommendation scores.
%
%
Following \SASRec, 
we employ the binary cross-entropy loss
to minimize the negative log-likelihood of correctly recommending the ground truth next item is as follows:
\begin{equation}
	\label{eqn:obj:primary}
	\min\limits_{\boldsymbol{\Theta}} -\sum \nolimits_{S_i \in T, v_j \not \in S_i} \log(\sigma(\hat{r}_{ik})) + \log(1-\sigma(\hat{r}_{ij})),
\end{equation}
where $T$ is the set of all the training sequences, 
$\boldsymbol{\Theta}$ is the set of learnable parameters (e.g., $V$ and $P$), 
$v_k$
is the ground-truth next item given $S_i$.
For each training sequence, we randomly generate one negative item $v_j$ for the optimization in each epoch.
All the learnable parameters are randomly initialized, 
and are optimized in an end-to-end manner.

\subsection{Complexity Analysis}
\label{sec:method:complexity}

%
As shown in the literature~\cite{sasrec}, 
the time complexity for each SA block 
in SA-based methods is $\mathcal{O} (n^2d+nd^2)$.
As a comparison, in \method, the time complexity of an attention block is only $\mathcal{O} (nd^2)$, which is theoretically better than that of the SA block.
Particularly, computing each SA block requires $6nd^2+2n^2d+2nd$ operations, 
while for our attention block, the number of computation operations is only $2nd^2+4d^2+2nd+2d$, which is also significantly
lower than that in SA blocks.
As will be shown in Section~\ref{sec:results:runtime}, 
the better time complexity could enable superior runtime performance of \method 
over the SA-based method \SASRec 
also on modern GPUs.
%

\section{Materials}
\label{sec:materials}

\subsection{Baseline Methods}
\label{sec:materials:baseline}

We compare \method with five state-of-the-art baseline methods:
	1) \Caser~\cite{tang2018personalized} learns users' preferences, 
	and the synergies among items using CNNs for sequential recommendation.
	2) \HAM~\cite{peng2021ham} explicitly models users' long-term preferences, the high-order and low-order association patterns among items, and the synergies among items to generate recommendations.
	3) \NARM~\cite{li2017neural} employs RNNs and attention mechanisms 
	to capture both users' short-term preferences and long-term preferences for the recommendation.
	4) \HGN~\cite{ma2019hierarchical} leverages attention mechanisms to identify
	important items and effective latent features for better recommendation.
	5) \SASRec~\cite{sasrec} stacks multiple SA blocks to learn users' preferences from the interacted items for sequential recommendation.
%
Note that \HGN and \SASRec have been compared with 
a comprehensive
set of other methods including \GRURec~\cite{hidasi2015session}, \GRURecP~\cite{hidasi2018recurrent} and
\NextItRec~\cite{yuan2018simple}, 
and have outperformed those methods. 
Therefore, we compare \method with \HGN and \SASRec instead of
the methods that they outperform.
For all the baseline methods, 
we use the implementations provided by the authors in
GitHub (Section~\ref{sec:reproducibility}).
%
%
%

\subsection{Datasets}
\label{sec:materials:datasets}

\begin{table*}
\footnotesize
  \caption{Dataset Statistics}
  \centering
  \label{tbl:dataset}
  \begin{threeparttable}
      \begin{tabular}{
	@{\hspace{15pt}}l@{\hspace{15pt}}
	@{\hspace{15pt}}r@{\hspace{15pt}}          
	@{\hspace{15pt}}r@{\hspace{15pt}}
	@{\hspace{15pt}}r@{\hspace{6pt}}
	@{\hspace{2pt}}r@{\hspace{15pt}}
        @{\hspace{15pt}}r@{\hspace{15pt}}
	}
        \toprule
        dataset      & \#users   & \#items &  \#intrns & {\#intrns/u} & \#u/i\\
        \midrule
        \Beauty       & 22,363  & 12,101  & 198,502    &  8.9  & 16.4 \\
        \Toys         & 19,412  & 11,924  & 167,597    &  8.6  & 14.1 \\
        \Children     & 48,296  & 32,871  & 2,784,423  & 57.6  & 84.7 \\
        \Comics       & 34,445  & 33,121  & 2,411,314  & 70.0  & 72.8 \\
        \MLOM         & 6,040   & 3,952   & 1,000,209  & 165.6 & 253.1\\
        \MLTM         & 129,780 & 13,663  & 9,926,480  & 76.5  & 726.5\\
        \bottomrule
      \end{tabular}
      \begin{tablenotes}[normal,flushleft]
      \begin{footnotesize}
      \item 
      In this table, ``\#users", ``\#items" and ``\#intrns" represents the number of 
      users, items and user-item interactions, respectively. 
      The column ``\#intrns/u" has the average 
      number of interactions of each user. 
      The column ``\#u/i" has the average number of interactions on each item.  
      \par
      \end{footnotesize}
      \end{tablenotes}
  \end{threeparttable}
  \vspace{-10pt}
\end{table*}

We evaluate the methods on six public benchmark datasets:
	1) Amazon-Beauty (\Beauty) and Amazon-Toys (\Toys) 
	are from Amazon reviews~\cite{amazon}.
	%
	These two datasets contain users' ratings and reviews 
	on beauty products and toys on Amazon, respectively.
	2) Goodreads-Children (\Children) and Goodreads-Comics
	(\Comics) are from the goodreads website~\cite{goodreads},
	%
	and have users' ratings and reviews on children and comic books, respectively.
	3) MovieLens-1M (\MLOM) and MovieLens-20M (\MLTM) are from the MovieLens website~\cite{movielens},
	and contain users' ratings on movies.
Following \SASRec, for \Beauty, \Toys and \MLOM, 
we only keep the users and items with at least 5 ratings.
For \Children, \Comics and \MLTM, which are not used in \SASRec, 
following \HAM, 
we keep users with at least 10 ratings, and items with at least 5 ratings. 
%
Following the literature~\cite{sasrec,peng2021ham}, 
we consider the ratings as users' implicit feedback, 
and convert ratings into binary values.
Particularly, for ratings with a range from 1 to 5, 
we convert ratings 4 and 5 to binary value 1, or 0 otherwise.
Table~\ref{tbl:dataset} presents the statistics of the six datasets after the processing.

\subsection{Experimental Protocol}
\label{sec:materials:protocl}

\subsubsection{Training, validation and testing sets}
\label{sec:materials:protocl:training}
 
Following \SASRec, on all the datasets, 
given the historical interactions $S$ of each user, 
we use the last item (i.e., $s_{|S|}$) in the history for testing, 
the second last item (i.e., $s_{|S|-1}$) for validation, 
and all the previous items for training. 
Also following \SASRec, 
for each training sequence $S$, we extract 
the last $n$ items $\{b_1, b_2, \dots, b_n\}$, 
and split it to $\{b_1, b_2\}$, $\{b_1, b_2, b_3\}$, 
$\cdots$, $\{b_1, b_2, \dots, b_n\}$.
We use all the resulted sequences 
(with necessary padding) for training.
We carefully tune the hyper parameters on the validation sets 
for \method and all the baseline methods 
using grid search, 
and use the best performing hyper parameters in terms of Recall@$10$ (Section~\ref{sec:materials:protocl:metric}) for testing.
To enable a fair comparison, we exhaustively search 
hyper parameters on a large parameter space. 
%
For reproducibility, we report the search range of hyper parameters, 
and the identified best performing hyper parameters 
for each method in Section~\ref{sec:reproducibility}.

\subsubsection{Evaluation metrics}
\label{sec:materials:protocl:metric}

Following the literature~\cite{sasrec,peng2021ham,ma2019hierarchical,tang2018personalized}, 
we use Recall@$k$ and NDCG@$k$ to evaluate \method and the baseline methods.
%
	 Recall@$k$ measures the proportion of sequences 
	in which the ground-truth next item (i.e., $s_{|S|-1}$ during validation and $s_{|S|}$ during testing) is correctly recommended.
	Particularly, for each sequence, the Recall@$k$ is $1$ 
	if the ground-truth next item is among the top-$k$ of the recommendation list, 
	or $0$ otherwise.
	NDCG@$k$ is the normalized discounted cumulative gain. 
	It is a widely used rank-aware metric to evaluate 
	the ranking quality of recommendation lists.
	Following the literature~\cite{sasrec,peng2021ham,ma2019hierarchical,tang2018personalized}, 
	in our experiments, the gain indicates whether the ground-truth next item is recommended (i.e., gain is $1$) or not (i.e., gain is $0$).	
%
In our experiments, 
we report the average results over all the users 
for each evaluation metric.
For all the evaluation metrics, 
a higher value indicates better performance.

\section{Experimental Results}
\label{sec:results}

\subsection{Overall Performance}
\label{sec:results:overall}

\begin{table*}[!t]
\small
  \caption{Overall Performance}
  \centering
  \label{tbl:overall_performance}
  \begin{footnotesize}
  \begin{threeparttable}
      \begin{tabular}{
		@{\hspace{0pt}}l@{\hspace{2pt}}
		@{\hspace{2pt}}l@{\hspace{2pt}}          
		@{\hspace{2pt}}r@{\hspace{2pt}}
		@{\hspace{2pt}}r@{\hspace{2pt}}
		@{\hspace{2pt}}r@{\hspace{2pt}}
		@{\hspace{2pt}}r@{\hspace{2pt}}
        @{\hspace{0pt}}r@{\hspace{2pt}}
        @{\hspace{2pt}}r@{\hspace{2pt}}
        @{\hspace{2pt}}r@{\hspace{2pt}}
        @{\hspace{2pt}}r@{\hspace{2pt}}
        @{\hspace{2pt}}l@{\hspace{2pt}}
        @{\hspace{2pt}}l@{\hspace{2pt}}
        @{\hspace{2pt}}r@{\hspace{2pt}}          
        @{\hspace{2pt}}r@{\hspace{2pt}}
        @{\hspace{2pt}}r@{\hspace{2pt}}
        @{\hspace{2pt}}r@{\hspace{2pt}}
        @{\hspace{0pt}}r@{\hspace{2pt}}
        @{\hspace{2pt}}r@{\hspace{2pt}}
        @{\hspace{2pt}}r@{\hspace{2pt}}
        @{\hspace{2pt}}r@{\hspace{0pt}}
	}
        \toprule
        & Dataset & \Caser & \HAM & \NARM & \HGN & \SASRec & \methodmu & \method & impv
        &&& \Caser & \HAM & \NARM & \HGN & \SASRec & \methodmu & \method & impv\\
        \midrule
	    \multirow{6}{*}{\rotatebox[origin=c]{90}{{\centering Recall@10}}}
        & \Beauty  & 0.0647 & \underline{0.0872} & 0.0554 & 0.0782 & 0.0777 & 0.0835 & \textbf{$\mathclap{^{\dagger~}}$0.0888} & 1.8\%
        && \multirow{6}{*}{\rotatebox[origin=c]{90}{{\centering NDCG@10}}}
        &  0.0371 & \underline{$\mathclap{^{\dagger~}}$0.0498} & 0.0291 & 0.0434 & 0.0417 & 0.0460 & \textbf{0.0494} & -0.8\%\\
        & \Toys    & 0.0675 & \underline{$\mathclap{^{\dagger~}}$0.1017} & 0.0557 & 0.0926 & 0.0938 & 0.0998 & \textbf{0.1005} & -1.2\%
        &&& 0.0396 & \underline{$\mathclap{^{\dagger~}}$0.0617} & 0.0308 & 0.0540 & 0.0525 & 0.0568 & \textbf{0.0593} & -3.9\%\\
        & \Children & 0.1303 & 0.1729 & 0.1126 & 0.1536 & \underline{0.1767} & 0.1718 & \textbf{$\mathclap{^{\dagger~}}$0.1906} & 7.9\%
        &&& 0.0764 & 0.1038 & 0.0585 & 0.0917 & \underline{0.1083} & 0.1042 & \textbf{$\mathclap{^{\dagger~}}$0.1158} & 6.9\%\\
        & \Comics  & 0.2320 & 0.3055 & 0.1304 & 0.2857 & \underline{0.3189} & 0.3169 & \textbf{$\mathclap{^{\dagger~}}$0.3286} & 3.0\%
        &&& 0.1642 & 0.2319 & 0.0720 & 0.2061 & \underline{0.2433} & 0.2392 & \textbf{$\mathclap{^{\dagger~}}$0.2445} & 0.5\%\\
        & \MLOM    & \underline{0.2874} & 0.2807 & 0.2349 & 0.2428 & 0.2589 & 0.2732 & \textbf{$\mathclap{^{\dagger~}}$0.2930} & 1.9\%
        &&& \underline{$\mathclap{^{\dagger~}}$0.1619} & 0.1598 & 0.1252 & 0.1374 & 0.1389 & 0.1481 & \textbf{0.1612} & -0.4\%\\
        & \MLTM    & 0.1739 & 0.1672 & OOM & 0.1588 & \underline{0.1892} & \textbf{$\mathclap{^{\dagger~}}$0.1921} & 0.1720 & 1.5\%
        &&& 0.0907 & 0.0895 & OOM & 0.0845 & \underline{0.0979} & \textbf{$\mathclap{^{\dagger~}}$0.0998} & 0.0882 & 1.9\%\\
        \midrule
        \multirow{6}{*}{\rotatebox[origin=c]{90}{{\centering Recall@20}}}
        & \Beauty  & 0.0869 & 0.1058 & 0.0835 & 0.0978 & \underline{0.1160} & 0.1226 & \textbf{$\mathclap{^{\dagger~}}$0.1291} & 11.3\%
        && \multirow{6}{*}{\rotatebox[origin=c]{90}{{\centering NDCG@20}}}
        &  0.0427 & \underline{0.0547} & 0.0362 & 0.0486 & 0.0514 & 0.0558 & \textbf{$\mathclap{^{\dagger~}}$0.0596} & 9.0\%\\
        & \Toys    & 0.0895 & 0.1182 & 0.0816 & 0.1095 & \underline{0.1292} & 0.1337 & \textbf{$\mathclap{^{\dagger~}}$0.1343} & 3.9\%
        &&& 0.0452 & \underline{0.0661} & 0.0373 & 0.0584 & 0.0614 & 0.0654 & \textbf{$\mathclap{^{\dagger~}}$0.0678} & 2.6\%\\
        & \Children & 0.1832 & 0.2083 & 0.1735 & 0.1854 & \underline{0.2362} & 0.2340 & \textbf{$\mathclap{^{\dagger~}}$0.2573} & 8.9\%
        &&& 0.0897 & 0.1131 & 0.0738 & 0.1001 & \underline{0.1233} & 0.1199 & \textbf{$\mathclap{^{\dagger~}}$0.1326} & 7.5\%\\
        & \Comics  & 0.2791 & 0.3316 & 0.1896 & 0.3122 & \underline{0.3650} & 0.3655 & \textbf{$\mathclap{^{\dagger~}}$0.3822} & 4.7\%
        &&& 0.1760 & 0.2388 & 0.0869 & 0.2131 & \underline{0.2549} & 0.2514 & \textbf{$\mathclap{^{\dagger~}}$0.2580} & 1.2\%\\
        & \MLOM    & \underline{0.3955} & 0.3324 & 0.3422 & 0.2971 & 0.3687 & 0.3891 & \textbf{$\mathclap{^{\dagger~}}$0.3990} & 0.9\%
        &&& \underline{$\mathclap{^{\dagger~}}$0.1892} & 0.1735 & 0.1524 & 0.1518 & 0.1665 & 0.1773 & \textbf{0.1880} & -0.6\%\\
        & \MLTM    & 0.2649 & 0.2129 & OOM & 0.2024 & \underline{0.2908} & \textbf{$\mathclap{^{\dagger~}}$0.2932} & 0.2664 & 0.8\%
        &&&  0.1136 & 0.1015 & OOM & 0.0960 & \underline{0.1234} & \textbf{$\mathclap{^{\dagger~}}$0.1252} & 0.1120 & 1.5\%\\
        \midrule
      \end{tabular}
           \begin{tablenotes}[normal,flushleft]
      \begin{footnotesize}
      \item
      For each dataset, the best performance among \method and its variant \methodmu is in \textbf{bold}, 
      the best performance among the baseline methods is \underline{underlined}, 
      and the overall best performance is indicated by a dagger (i.e.,  $\dagger$). 
      The column ”impv” presents the percentage improvement of the best performing variant of \method (\textbf{bold}) 
      over the best performing baseline methods (\underline{underlined}).
      ``OOM" represents the out-of-memory issue. 
      \par
      \end{footnotesize}
      \end{tablenotes}
  \end{threeparttable}
  \end{footnotesize}
\end{table*}

Table~\ref{tbl:overall_performance} presents the overall performance of \method, 
its variant \methodmu
and the state-of-the-art baseline methods at Recall@$k$ and NDCG@$k$ on six benchmark datasets.
In \methodmu, we remove user embeddings (i.e., $\mathbf{u}_i$) when calculating recommendation scores (Equation~\ref{eqn:score}).
%
%
For \NARM, on \MLTM, with the implementation provided by the authors, 
we get the out-of-memory (OOM) issue 
on NVIDIA Volta V100 GPUs with 16 GB memory.
%

As shown in Table~\ref{tbl:overall_performance}, overall, 
\method is the best performing method on the six datasets.
In terms of Recall@$10$, \method achieves the best performance on four out of six datasets (i.e., \Beauty, \Children, \Comics and \MLOM), and the second best performance on \Toys.
Similarly, at Recall@$20$, \method also significantly outperforms all the baseline methods on five out of six datasets except \MLTM.
%
%
We observe a similar trend at NDCG@$k$.
In terms of NDCG@$10$ and NDCG@$20$, \method achieves the best or second best performance on five out of six datasets except for \MLTM.
%
%
We notice that on \MLTM, compared to the best performing baseline method, the performance of \method is considerably worse.
%
%
However, 
without user embeddings, \methodmu could still 
achieve the best performance on \MLTM.
These results demonstrate the strong performance of \method and its variant \methodmu
on different recommendation datasets. 
It is also worth noting that, different from \method, 
\Caser and \HAM do not employ attention mechanisms in learning users' preferences.
As shown in Table~\ref{tbl:overall_performance}, 
compared to \Caser, 
\method achieves superior performance on five out of 
six datasets at Recall@$10$, and all the six datasets at Recall@$20$.
Similarly, \method also outperforms \HAM on five datasets 
in terms of Recall@$10$ and all the six datasets in terms of Recall@$20$.
These results demonstrate the effectiveness of attention mechanisms 
for sequential recommendation

\subsubsection{Comparison between \method and \SASRec}
\label{sec:results:overall:sasrec}

The SA-based method \SASRec learns user preferences 
using stacked SA blocks, and has been demonstrated 
the state-of-the-art performance in the literature~\cite{peng2021ham,sasrec}.
However, as shown in Table~\ref{tbl:overall_performance}, \method consistently outperforms \SASRec on five out of six datasets (i.e., \Beauty, \Toys, \Children, \Comics and \MLOM).
Particularly, compared to \SASRec, in terms of Recall@$10$, \method achieves a significant improvement of 9.1\%, on average, over the five datasets.
In terms of NDCG@$10$, \method also significantly outperforms \SASRec with an average improvement of 11.0\% over the five datasets.
On \MLTM, although the performance of \method is worse than that of \SASRec, 
\methodmu could still outperform \SASRec by a considerable margin.
\SASRec recursively aggregates and updates item representations 
through stacked SA blocks 
to capture both users' short-term and long-term preferences.
However, as discussed in Section~\ref{sec:motivation}, 
the learned attention maps in \SASRec 
could suffer from the localization-deficit issue, and thus, in each block, 
updating the item representations based on the attention map could 
induce a dramatic change in the item representations over blocks, and eventually impair the recommendation performance.
%
%
%
Differently from \SASRec, in \method, 
we reuse item representations through blocks to mitigate the issue,
and also  
explicitly learn user embeddings to capture users' long-term preferences.
%
Therefore, \method could enable substantially 
better recommendation performance over \SASRec on the benchmark datasets.
%
%
%
%

\subsubsection{Comparison between \method and \HGN}
\label{sec:results:overall:hgn}

Table~\ref{tbl:overall_performance} also shows that overall 
\method substantially outperforms \HGN on all the six benchmark datasets.
For example, in terms of Recall@$10$ and Recall@$20$, 
\method achieves significant improvement over \HGN on all the six datasets.
On average, compared to \HGN, \method achieves a substantial improvement of 15.0\% and 30.3\% at Recall@$10$ and Recall@$20$, respectively.
Similarly to  \method, \HGN also learns attention weights to aggregate items 
and capture users' preferences.
However, \HGN directly learns users' preferences using one attention layer, 
while \method stacks multiple attention-blocks 
to model users' preferences recursively.
Given the complex nature of users' preferences~\cite{yakhchi2021learning}, one layer may not be sufficient to fully capture users' preferences.
Therefore, by recursively modeling users' preferences via multiple blocks, \method could achieve superior recommendation performance over \HGN on all the six benchmark datasets.
Note that in \HGN, there is no mechanism to stack multiple layers, 
and thus, it could be non-trivial to extend \HGN to a multi-layer version.

\subsubsection{Comparison between \method and \NARM}
\label{sec:results:overall:narm}

Table~\ref{tbl:overall_performance} shows that compared to the RNNs-based method \NARM, \method demonstrates superior performance on all the six datasets at all the evaluation metrics.
\NARM primarily uses RNNs to recurrently learn users' preferences.
As shown in the literature~\cite{vaswani2017attention}, due to the recurrent nature, RNNs may not be effective in modeling long-range dependencies in the sequence.
In \method, similarly to that in SA-based methods, 
we use attention mechanisms to model users' preferences, 
which could be more effective than RNNs in capturing the long-range dependencies as demonstrated in the literature~\cite{vaswani2017attention}.
Therefore \method could significantly outperform 
RNNs-based method \NARM on benchmark datasets. 




\subsection{User Embedding Analysis}
\label{sec:results:user}

We conduct an analysis to verify the importance of learning user embeddings in \method.
Specifically,  
we compare the performance of \method 
and its variant \methodmu 
to verify the effectiveness of user embeddings.
We also investigate the similarities between the 
representation of users' short-term preferences and the user embeddings to verify if 
they capture different preferences of users.

\subsubsection{Effectiveness of user embeddings}
\label{sec:results:user:effect}

%
%
%
%

As shown in Table~\ref{tbl:overall_performance}, 
without user embeddings, the performance of \methodmu drops significantly on five out six datasets except for \MLTM.
For example, in terms of Recall@10, 
on \Children and \Comics, \methodmu underperforms \method at 9.9\% and 3.6\%, respectively.
On average. over the five datasets, \methodmu underperforms \method at 5.3\% in Recall@10.
In \method, we hypothesize that the attention blocks may not be able to effectively capture users' long-term preferences. 
Therefore, following the literature~\cite{peng2021ham,ma2019hierarchical}, we explicitly learn user embeddings to 
better capture the long-term preferences in the model. 
These results demonstrate the importance of separately modeling users' short-term preferences and long-term preferences for sequential recommendation.
We notice that on \MLTM, different from that in the other datasets, 
including user embeddings results in worse performance.
This might be due to the reason that in some datasets, 
users' interactions primarily depend on their short-term preferences~\cite{stamp}, and thus, 
incorporating the long-term preferences when generating the recommendations 
may not benefit the performance.
 
\subsubsection{Similarities between the representation of users' short-term preferences 
	and user embeddings}
\label{sec:results:user:sim}

In \method, as discussed in Section~\ref{sec:method:long}, we view the outputs from the attention blocks 
(i.e., $\mathbf{h}_i^{(n_b)}$) as the 
representations of users' short-term preferences, 
and combine them with the user embeddings (i.e., $\mathbf{u}_i$) 
for the recommendation.
Intuitively, 
users' short-term and long-term preferences might be different,
and thus, combining both $\mathbf{h}_i^{(n_b)}$ and $\mathbf{u}_i$ 
could enable better performance.
%
To verify this, 
we calculate the cosine similarities between $\mathbf{h}_i^{(n_b)}$ and $\mathbf{u}_i$. 
%
%
%
For each user $u_i$, we also randomly sample another user $u_j$ and 
calculate the cosine similarity between $\mathbf{h}_i^{(n_b)}$ and $\mathbf{u}_j$. 
Generally, the short-term preferences and long-term preferences from two different users 
should be considerably different, 
and can be used as a baseline to evaluate those from the same user.
%

\begin{figure}[!h]
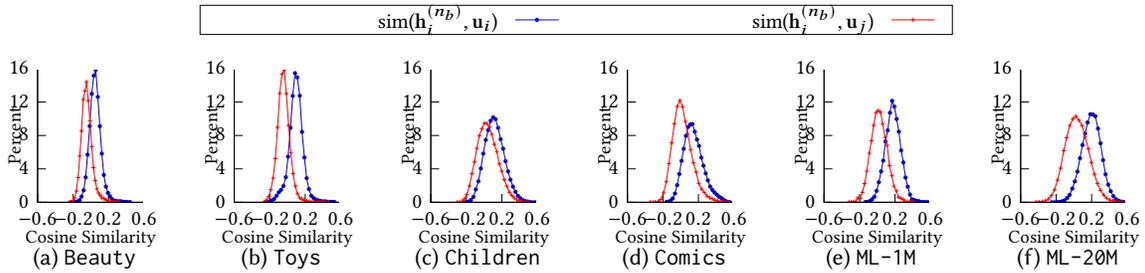

	\centering
	\footnotesize
	
	\begin{subfigure}{\linewidth}
		\centering
		\input{plots/sim/legendsim.tex}
		\vspace*{-10pt}
	\end{subfigure}	
	\begin{subfigure}{0.16\linewidth}
		\centering
		\input{plots/sim/Beauty.tex}
		\vspace*{10pt}
		\caption{\Beauty}
		\label{fig:sim:beauty}
	\end{subfigure}
	\begin{subfigure}{0.16\linewidth}
		\centering
		\input{plots/sim/Toys.tex}
		\vspace*{10pt}
		\caption{\Toys}
		\label{fig:sim:toys}
	\end{subfigure}
	\begin{subfigure}{0.16\linewidth}
		\centering
		\input{plots/sim/Children.tex}
		\vspace*{10pt}                    
		\caption{\Children}
		\label{fig:sim:child}
	\end{subfigure}
	\begin{subfigure}{0.16\linewidth}
		\centering
		\input{plots/sim/Comics.tex}
		\vspace*{10pt}
		\caption{\Comics}
		\label{fig:sim:comics}
	\end{subfigure}
	\begin{subfigure}{0.16\linewidth}
		\centering
		\input{plots/sim/ml1m.tex}
		\vspace*{10pt}
		\caption{\MLOM}
		\label{fig:sim:ml1m}
	\end{subfigure}
	\begin{subfigure}{0.16\linewidth}
		\centering
		\input{plots/sim/ml20m.tex}
		\vspace*{10pt}
		\caption{\MLTM}
		\label{fig:sim:ml20m}
	\end{subfigure}
	\vspace{-10pt}
	\caption{Cosine Similarities Between the representation of users' short-term and long-term preferences}
	\label{fig:sim}
	\vspace{-10pt}
\end{figure}

Figure~\ref{fig:sim} presents the distribution of the cosine similarities
between the preference representations
from the same user (i.e., $\mathbf{h}_i^{(n_b)}$ and $\mathbf{u}_i$), 
and different users (i.e., $\mathbf{h}_i^{(n_b)}$ and $\mathbf{u}_j$) 
on the six benchmark datasets.
%
%
As shown in Figure~\ref{fig:sim}, on all the datasets, 
%
the distribution 
of the similarities between $\mathbf{h}_i^{(n_b)}$ and $\mathbf{u}_i$ 
is close to that between $\mathbf{h}_i^{(n_b)}$ and $\mathbf{u}_j$.
Recall that since from different users, 
$\mathbf{h}_i^{(n_b)}$ and $\mathbf{u}_j$ should generally represent different preferences.
Therefore, these results reveal that $\mathbf{h}_i^{(n_b)}$ and $\mathbf{u}_i$ capture different preferences of users,
and thus, as shown in Table~\ref{tbl:overall_performance}, 
using both of them for the recommendation could significantly improve the performance on most of the datasets. 
%
%
%
%
%

\subsection{Stability Analysis}
\label{sec:results:stability}

We also analyze to evaluate the stability 
of \SASRec and \method on 
learning deep 
and wide models.
Following the literature~\cite{cheng2016wide}, 
we use the number of blocks (i.e., $n_b$) and 
the embedding dimensions (i.e., $d$) 
to represent the depth and width of models, respectively.
%
%
To enable a fair comparison, in this analysis, we also compare \SASRec with \methodmu to eliminate the effects of user embeddings.
%
We use the identified best-performing hyper parameters on \SASRec 
for \SASRec, \methodmu and \method, 
and study how $d$ and $n_b$ affect the performance.

\subsubsection{Stability over the number of blocks}
\label{sec:results:stability:layer}

\begin{figure}[!h]
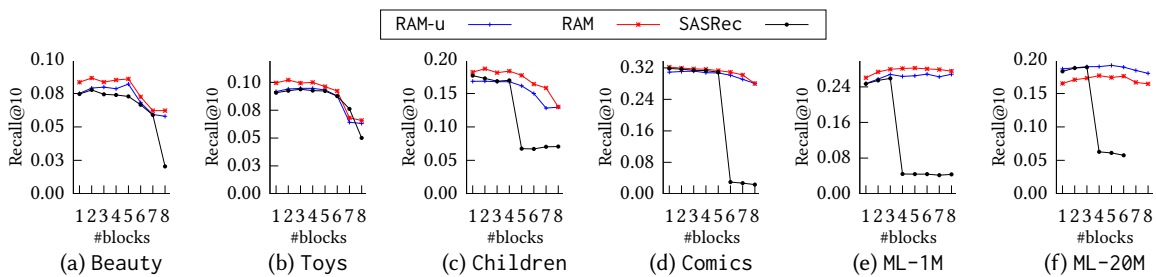

	\centering
	\footnotesize
	\begin{subfigure}{\linewidth}
		\centering
		\input{plots/stability/legend.tex}
		\vspace*{-15pt}
	\end{subfigure}
	\begin{subfigure}{0.16\linewidth}
		\centering
		\input{plots/stability/Beauty_sta.tex}
		\vspace*{15pt}
		\caption{\Beauty}
		\label{fig:sta:beauty}
	\end{subfigure}
	\begin{subfigure}{0.16\linewidth}
		\centering
		\input{plots/stability/Toys_sta.tex}
		\vspace*{15pt}
		\caption{\Toys}
		\label{fig:sta:toys}
	\end{subfigure}
	\begin{subfigure}{0.16\linewidth}
		\centering
		\input{plots/stability/Children_sta.tex}
		\vspace*{15pt}                    
		\caption{\Children}
		\label{fig:sta:child}
	\end{subfigure}
	\begin{subfigure}{0.16\linewidth}
		\centering
		\input{plots/stability/Comics_sta.tex}
		\vspace*{15pt}
		\caption{\Comics}
		\label{fig:sta:comics}
	\end{subfigure}
	\begin{subfigure}{0.16\linewidth}
		\centering
		\input{plots/stability/ml1m_sta.tex}
		\vspace*{15pt}
		\caption{\MLOM}
		\label{fig:sta:ml1m}
	\end{subfigure}
	\begin{subfigure}{0.16\linewidth}
		\centering
		\input{plots/stability/ml20m_sta.tex}
		\vspace*{15pt}
		\caption{\MLTM}
		\label{fig:sta:ml20m}
	\end{subfigure}
	%
	\vspace{-10pt}
	\caption{Performance over a different number of blocks}
	\label{fig:sta}
	\vspace{-10pt}
\end{figure}

Figure~\ref{fig:sta} presents the performance of \SASRec, \methodmu and \method 
over the different number of blocks on the six benchmark datasets.
%
On \MLTM, we have the out-of-memory issue on \SASRec when $n_b > 6$.
Thus, on \MLTM, we only report the performance of \SASRec when $n_b \leq 6$.
%
%
%
%
As shown in Figure~\ref{fig:sta}, 
both \methodmu and \method are more stable than \SASRec on learning deep models.
Particularly, for \methodmu and \method, on all the datasets, 
they could perform reasonably well when the model is very deep (e.g., $n_b=8$).
%
%
However, for \SASRec, 
on \Beauty and \Toys, the performance drops dramatically 
when the number of blocks (i.e., $n_b$) is larger than seven.
In \Children and \Comics, the performance also degrades dramatically 
when $n_b>4$ and $n_b>5$, respectively.
Moreover, on \MLOM and \MLTM, \SASRec performs very poorly when $n_b > 3$.
As discussed in Section~\ref{sec:method:attention:attention}, 
the attention maps learned in \SASRec could suffer from the localization-deficit issue.
As a consequence, in the lower blocks,
the item representations 
updated with the localization-deficit attention maps could 
change dramatically, 
and further impair the localization of the following maps.  
%
Through multiple blocks, this problem could be amplified, and thus, \SASRec 
could be unstable on learning deep models.
Different from \SASRec, in \methodmu and \method, 
we do not update the item representations, and 
reuse them through all the blocks to avoid this issue.
%
%
%
%
Therefore, \methodmu and \method could learn deeper models than 
\SASRec for better performance.
We notice that on \Toys, the performance of 
\methodmu and \method drop considerably when $n_b>6$.
As shown in Table~\ref{tbl:dataset}, 
\Toys is the most sparse dataset in our experiments.
As a result, 
the sparse interactions in \Toys may not enable learning very deep models. 
%
However, on \Toys, \methodmu and \method could still outperform \SASRec when $n_b=8$. 
%
It is also worth noting that, 
as will be shown in Section~\ref{sec:reproducibility},
on \Toys, \MLOM and \MLTM, 
the best performing \methodmu and \method have at least 4 blocks,
%
%
%
which reveals that on specific datasets, learning moderately 
deep models could benefit the performance.

%

We also investigate the output (i.e., $\mathbf{h}^{(m)}$) from each attention block 
of \SASRec and \method to better understand why \method 
could be more stable than \SASRec in learning deep models.
Intuitively, to ensure stability, 
the outputs of consecutive attention blocks should not change significantly.
To verify this, we calculate the cosine similarity 
between the outputs of consecutive attention blocks 
(i.e., sim($h^{(m)}$, $h^{(m+1)}$)).
Specifically, 
we use the best-performing hyper parameters (Section~\ref{sec:reproducibility}) 
for \SASRec and \method, 
and export the outputs over attention blocks to calculate their similarities.
We find that the cosine similarity 
between $h^{(m)}$ and $h^{(m+1)}$ in \method 
is considerably higher than that in \SASRec.
For example, on \MLOM, in \method, 
sim($h^{(1)}$, $h^{(2)}$), and sim($h^{(2)}$, $h^{(3)}$) 
is $0.819$ and $0.797$, respectively.
However, for \SASRec, sim($h^{(1)}$, $h^{(2)}$) and sim($h^{(2)}$, $h^{(3)}$) are both 
$0.719$. 
The higher similarities in \method over \SASRec 
indicate that in \method, the outputs over blocks are more consistent 
compared to those in \SASRec, and thus, could enable better stability.
%


\subsubsection{Stability over embedding dimensions}
\label{sec:results:stability:dim}

\begin{figure}[!h]
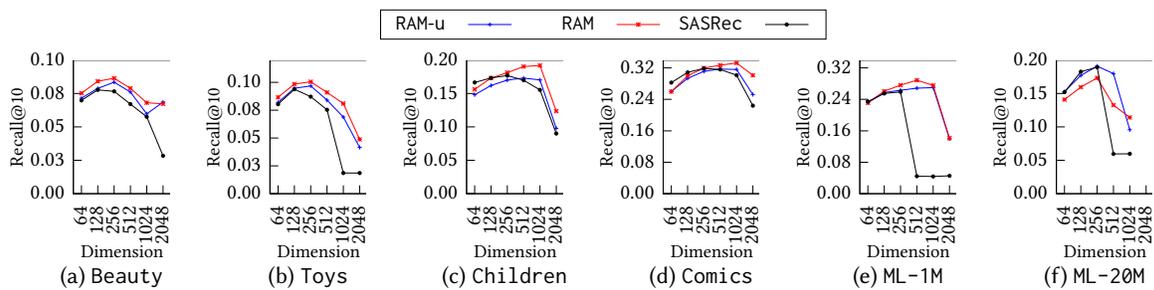

	\vspace{0pt}
	\centering
	\footnotesize
	\begin{subfigure}{\linewidth}
		\centering
		\input{plots/stability/legend.tex}
		\vspace*{-15pt}
	\end{subfigure}
	\begin{subfigure}{0.16\linewidth}
		\centering
		\input{./plots/stabilityDims/Beauty_staD.tex}
		\vspace*{20pt}                    
		\caption{\Beauty}
		\label{fig:stad:beauty}
	\end{subfigure}
	\begin{subfigure}{0.16\linewidth}
		\centering
		\input{plots/stabilityDims/Toys_staD.tex}
		\vspace*{20pt}                    
		\caption{\Toys}
		\label{fig:stad:toys}
	\end{subfigure}
	\begin{subfigure}{0.16\linewidth}
		\centering
		\input{plots/stabilityDims/Children_staD.tex}
		\vspace*{20pt}                    
		\caption{\Children}
		\label{fig:stad:child}
	\end{subfigure}
	\begin{subfigure}{0.16\linewidth}
		\centering
		\input{plots/stabilityDims/Comics_staD.tex}
		\vspace*{20pt}
		\caption{\Comics}
		\label{fig:stad:comics}
	\end{subfigure}
	\begin{subfigure}{0.16\linewidth}
		\centering
		\input{plots/stabilityDims/ml1m_staD.tex}
		\vspace*{20pt}
		\caption{\MLOM}
		\label{fig:stad:ml1m}
	\end{subfigure}
	\begin{subfigure}{0.16\linewidth}
		\centering
		\input{plots/stabilityDims/ml20m_staD.tex}
		\vspace*{20pt}
		\caption{\MLTM}
		\label{fig:stad:ml20m}
	\end{subfigure}
	\vspace{-10pt}
	\caption{Performance over different embedding dimensions}
	\label{fig:stad}
	\vspace{-10pt}
\end{figure}

Figure~\ref{fig:stad} presents the performance 
of \SASRec, \methodmu and \method over different embedding dimensions 
on the six datasets.
%
%
On \MLTM, when $d=2048$, for all the three methods, 
we cannot finish the training within 48 hours.
Thus, on \MLTM, we only report the results when $d<2048$. 
%
%
%
As shown in Figure~\ref{fig:stad}, overall, 
\methodmu and \method are also more stable than \SASRec on learning wide models.
On all the datasets, 
both \methodmu and \method could perform reasonably well when the model is very wide  
(i.e., $d$ is $1024$ or $2048$).
However, for \SASRec, on \Beauty and \Toys, 
the performance drops dramatically when the embedding dimension is larger than 
$1024$ and $512$, respectively.
Similarly, on \MLOM and \MLTM, 
\SASRec also performs poorly when $d>256$.
These results demonstrate that 
reusing item representations could also enable \methodmu 
and \method to learn wider models than \SASRec for better performance.
We notice that on \Children and \Comics, 
\SASRec could achieve reasonable performance when $d$ is very large (i.e., $2048$).
However, on these datasets, 
when $d=2048$, 
\methodmu and \method could 
still outperform \SASRec by a considerable margin, indicating the capability of \methodmu and \method 
in learning wide models. 
%
%
%
%
%
%
%


\subsection{Run-time Performance over the Number of Attention Blocks}
\label{sec:results:runtime}



\begin{figure}[!h]
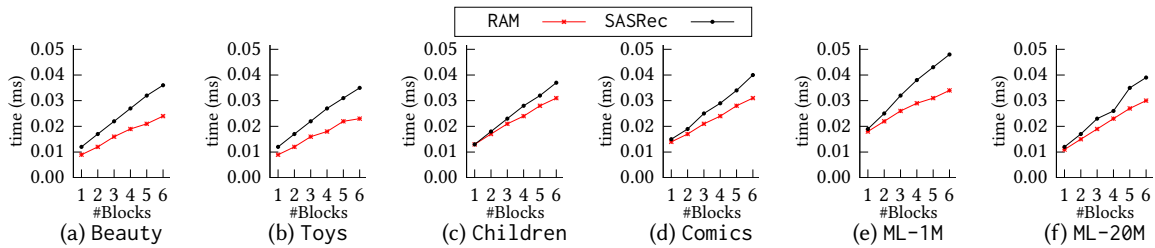

	\vspace{-5pt}
	\centering
	\footnotesize
	\begin{subfigure}{\linewidth}
		\centering
		\input{plots/runtimeLayers/legend2.tex}
		\vspace*{-20pt}
	\end{subfigure}	
	\begin{subfigure}{0.16\linewidth}
		\centering
		\input{plots/runtimeLayers/Beauty_layers.tex}
		\vspace*{10pt}
		\caption{\Beauty}
		\label{fig:runtime:beauty}
	\end{subfigure}
	\begin{subfigure}{0.16\linewidth}
		\centering
		\input{plots/runtimeLayers/Toys_layers.tex}
		\vspace*{10pt}
		\caption{\Toys}
		\label{fig:runtime:toys}
	\end{subfigure}
	\begin{subfigure}{0.16\linewidth}
		\centering
		\input{plots/runtimeLayers/Children_layers.tex}
		\vspace*{10pt}
		\caption{\Children}
		\label{fig:runtime:child}
	\end{subfigure}
	\begin{subfigure}{0.16\linewidth}
		\centering
		\input{plots/runtimeLayers/Comics_layers.tex}
		\vspace*{10pt}
		\caption{\Comics}
		\label{fig:runtime:comics}
	\end{subfigure}
	\begin{subfigure}{0.16\linewidth}
		\centering
		\input{plots/runtimeLayers/ml1m_layers.tex}
		\vspace*{10pt}                    
		\caption{\MLOM}
		\label{fig:runtime:ml1m}
	\end{subfigure}
	\begin{subfigure}{0.16\linewidth}
		\centering
		\input{plots/runtimeLayers/ml20m_layers.tex}
		\vspace*{10pt}                    
		\caption{\MLTM}
		\label{fig:runtime:ml20m}
	\end{subfigure}
\vspace{-10pt}
\caption{Runtime Performance over Different Number of Layers}
\label{fig:runtime}
\vspace{-10pt}
\end{figure}

We conduct an analysis to evaluate the run-time performance of \method and \SASRec during testing. 
Similarly to that in Section~\ref{sec:results:stability:layer}, 
we apply the best performing hyper parameters on \SASRec for \SASRec and \method, 
and study their run-time performance during testing over the different number of attention blocks.
To enable a fair comparison, 
we perform the evaluation for both \SASRec and \method 
using NVIDIA Volta V100 GPUs, and report the 
average computation time per user over five runs 
in Figure~\ref{fig:runtime}.
We focus on the run-time performance during 
testing due to the fact that it
could signify the models' latency in real-time recommendation, 
that could significantly affect the user experience and thus revenue.

As shown in Figure~\ref{fig:runtime}, 
on all the datasets, 
\method has better run-time performance than \SASRec over all the blocks, 
and the improvement increases as the number of blocks increases.
Particularly, 
when $n_b$ is the best performing one for \SASRec 
on each dataset (e.g., $n_b=2$ on \Beauty as in Section~\ref{sec:reproducibility}), 
in terms of the run-time performance, 
\method could achieve 
a 16.6\% speedup over \SASRec on the six datasets.
As shown in Figure~\ref{fig:sta}, 
with the same hyper parameters, on all the datasets except \MLTM, 
\method could also outperform \SASRec in terms of 
the recommendation performance.
These results demonstrate that compared to \SASRec, 
\method could generate more satisfactory recommendations in lower latency, 
and thus, could significantly improve the user experience.
Note that on GPUs, all the computations are performed parallelly.
However, when calculating the time complexity (Section~\ref{sec:method:complexity}), 
we assume the computations are serial.
%
Therefore, in terms of the run-time performance on modern GPUs, 
the improvement of \method over \SASRec may 
not be as significant as that 
on theoretical time complexity.  
However, in many applications, the recommendation model could be 
deployed on edge devices with limited computing resources. 
%
In these applications, as suggested by the time complexity comparison (Section~\ref{sec:method:complexity}), 
\method could achieve more substantial speedup over \SASRec.

\section{Reproducibility}
\label{sec:reproducibility}

\begin{table*}[!t]
\footnotesize
  \caption{\mbox{Hyper Parameters for \method, \methodmu and Baseline Methods}}
  \centering
  \label{tbl:para}
  \begin{threeparttable}
      \begin{tabular}{
        @{\hspace{2pt}}l@{\hspace{2pt}}
        @{\hspace{1pt}}r@{\hspace{1pt}}
        @{\hspace{1pt}}r@{\hspace{1pt}}
        @{\hspace{1pt}}r@{\hspace{1pt}}
        @{\hspace{1pt}}r@{\hspace{1pt}}
        @{\hspace{2pt}}r@{\hspace{1pt}}
	@{\hspace{2pt}}c@{\hspace{2pt}}
        @{\hspace{1pt}}r@{\hspace{2pt}}
        @{\hspace{1pt}}r@{\hspace{1pt}}
        @{\hspace{1pt}}r@{\hspace{1pt}}
        @{\hspace{1pt}}r@{\hspace{1pt}}
        @{\hspace{1pt}}r@{\hspace{1pt}}
        @{\hspace{1pt}}r@{\hspace{1pt}}
        @{\hspace{2pt}}c@{\hspace{2pt}}
        @{\hspace{1pt}}r@{\hspace{2pt}}
        @{\hspace{1pt}}r@{\hspace{1pt}}
        @{\hspace{2pt}}c@{\hspace{2pt}}
        @{\hspace{1pt}}r@{\hspace{2pt}}
        @{\hspace{1pt}}r@{\hspace{1pt}}
        @{\hspace{1pt}}r@{\hspace{1pt}}
        @{\hspace{1pt}}r@{\hspace{1pt}}
        @{\hspace{2pt}}c@{\hspace{2pt}}
        @{\hspace{1pt}}r@{\hspace{2pt}}
        @{\hspace{1pt}}r@{\hspace{1pt}}
        @{\hspace{1pt}}r@{\hspace{1pt}}
        @{\hspace{1pt}}r@{\hspace{1pt}}
        @{\hspace{2pt}}c@{\hspace{2pt}}
        @{\hspace{1pt}}r@{\hspace{2pt}}
        @{\hspace{1pt}}r@{\hspace{1pt}}
        @{\hspace{1pt}}r@{\hspace{1pt}}
        @{\hspace{1pt}}r@{\hspace{1pt}}
        @{\hspace{2pt}}c@{\hspace{2pt}}
        @{\hspace{1pt}}r@{\hspace{2pt}}
        @{\hspace{1pt}}r@{\hspace{1pt}}
        @{\hspace{1pt}}r@{\hspace{1pt}}
        @{\hspace{2pt}}r@{\hspace{2pt}}
        }
        \toprule
        \multirow{2}{*}{Dataset} & \multicolumn{5}{c}{\Caser} && \multicolumn{6}{c}{\HAM} 
        && \multicolumn{2}{c}{\NARM} && \multicolumn{4}{c}{\HGN} && \multicolumn{4}{c}{\SASRec} &&
        \multicolumn{4}{c}{\methodmu} && \multicolumn{4}{c}{\method}\\
        \cmidrule(lr){2-6} \cmidrule(lr){8-13} \cmidrule(lr){15-16} \cmidrule(lr){18-21} \cmidrule(lr){23-26}
        \cmidrule(lr){28-31} \cmidrule(lr){33-36}
	&  $d$ & $n_s$ & $n_p$ & $n_v$ & $n_f$
	&& $d$ & $n_s$ & $n_p$ & $\lambda$ & $n_l$ & $n_o$
        && $d$ & $l_r$
        && $d$ & $n_s$ & $n_p$ & $\lambda$
        && $d$ & $n$ & $n_h$ & $n_b$
        && $d$ & $n$ & $n_h$ & $n_b$
        && $d$ & $n$ & $n_h$ & $n_b$\\
        \midrule
        \Beauty   & 512 & 4 & 1 & 2 & 8 && 512 & 3 & 2 & 1e-3 & 1 & 1 && 512 & 1e-3 && 512 & 3 & 2 & 1e-3 && 128 & 75  & 4 & 2 && 256 & 75  & 4 & 2 && 256 & 75 & 16 & 3\\ 
	\Toys	  & 512 & 4 & 1 & 1 & 4 && 512 & 3 & 1 & 1e-3 & 1 & 1 && 512 & 1e-3 && 512 & 3 & 1 & 1e-4 && 128 & 50  & 2 & 3 && 256 & 50  & 4 & 4 && 256 & 50 & 8 & 4\\
	\Children & 512 & 4 & 1 & 2 & 4 && 256 & 4 & 2 & 1e-4 & 1 & 2 && 256 & 1e-4 && 256 & 3 & 1 & 0    && 256 & 175 & 2 & 1 && 256 & 200 & 1 & 2 && 512 & 100 & 1 & 1\\
	\Comics   & 512 & 4 & 1 & 1 & 4 && 512 & 3 & 1 & 0    & 1 & 3 && 256 & 1e-3 && 512 & 3 & 1 & 0    && 256 & 200 & 1 & 1 && 512 & 200 & 1 & 1 && 512 & 200 & 1 & 1\\
	\MLOM     & 128 & 6 & 1 & 4 & 16&& 512 & 5 & 1 & 1e-3 & 2 & 1 && 512 & 1e-4 && 128 & 4 & 1 & 1e-3 && 256 & 150 & 4 & 3 && 512 & 200 & 2 & 5 && 512 & 200 & 4 & 4\\
	\MLTM     & 256 & 6 & 1 & 2 & 8 && 256 & 5 & 2 & 1e-3 & 3 & 3 && OOM & OOM  && 128 & 4 & 2 & 1e-3 && 256 & 150 & 2 & 3 && 512 & 150 & 8 & 5 && 256 & 100 & 4 & 4\\
        \bottomrule
      \end{tabular}
      \begin{tablenotes}[normal,flushleft]
      \begin{scriptsize}
      \item
      This table presents the best performing hyper parameters of all the methods on the six benchmark datasets.
      In the table. ``OOM" represents the out of memory issue.
 	%
          %
          \par
      \end{scriptsize}
      \end{tablenotes}
  \end{threeparttable}
\end{table*}

We implement \method in python 3.7.11 with PyTorch 1.10.2~\footnote{\url{https://pytorch.org}}.
We use Adam optimizer with learning rate 1e-3 for \method on all the datasets.
The source code and processed data are available on Google Drive~\footnote{\mbox{\url{https://drive.google.com/drive/folders/1JxJ_-oE7B0I9mu39rLJ--HE5bh8c7Dvb?usp=sharing}}}. 
For \SASRec, \methodmu and \method, 
we search the embedding dimension $d$ in $\{64, 128, 256, 512\}$, 
the length of the fixed-length sequences $n$ in $\{50, 75, 100, 125, 150, 175, 200\}$, 
the number of heads $n_h$ in $\{1, 2, 4, 8, 16\}$, 
and the number of blocks $n_b$ in $\{1, 2, 3, 4, 5\}$.
We use GELU~\cite{hendrycks2016gaussian} as 
the activation function in the feed-forward layer for 
all the three methods.
We use the PyTorch implementation in GitHub~\cite{sasrec_pytorch}
for \SASRec.
%
%
%
%
For \HGN~\cite{hgn_pytorch},
we search $d$ in $\{64, 128, 256, 512\}$, 
the length of the subsequences $n_s$ in $\{3, 4, 5\}$, 
the number of items $n_p$ to calculate recommendation errors 
during training in $\{1, 2\}$, 
and the regularization factor $\lambda$ in \{0, 1e-3, 1e-4\}.
%
%
For \NARM~\cite{narm_pytorch},
we search $d$ in $\{64 ,128, 256, 512\}$ and 
the learning rate $l_r$ in \{1e-2, 1e-3, 1e-4\}. 
For \HAM~\cite{ham_pytorch},
we search $d$ in $\{64, 128, 256, 512\}$, 
$n_s$ in $\{3, 4, 5\}$,  
$n_p$ in $\{1, 2\}$, 
$\lambda$ in \{0, 1e-3, 1e-4\}, 
the number of items in low order $n_l$ in $\{1, 2, 3\}$,
and the order of item synergies $n_o$ in $\{1, 2, 3\}$. 
For \Caser~\cite{caser_pytorch},
we search $d$ in $\{64, 128, 256, 512\}$,
$n_s$ in $\{4, 5, 6\}$,  
$n_p$ in $\{1, 2\}$, 
the number of vertical filters in CNNs $n_v$ in $\{1, 2, 4\}$, 
and the number of horizontal filters in CNNs $n_f$ in $\{4, 8, 16\}$.
We report the best performing hyper parameters from the validation set in Table~\ref{tbl:para}.






\section{Conclusion}
\label{sec:conclusion}

In this manuscript, we identified the localization-deficit 
issue in SA-based sequential recommendation methods.
%
%
%
To mitigate the effects of this issue and 
improve the recommendation performance, 
we reused item representations throughout all the attention blocks
and developed the novel method \method for sequential recommendation.
\method models users' short-term preferences using the attention blocks with feed-forward layers, 
and also users' long-term preferences explicitly using a user embedding.  
We also developed a variant of \method with the user embedding removed. 
We extensively evaluated \method and \methodmu against 
five state-of-the-art baseline methods \Caser, \HAM, \NARM, \HGN and \SASRec on six benchmark datasets
\Beauty, \Toys, \Children, \Comics, \MLOM and \MLTM.
Our experimental results show that \method and \methodmu can achieve superior performance 
over the baseline methods on benchmark datasets, with an improvement of up to 11.3\%.
Our experiments also show that explicitly learning and integrating users' long-term preferences can benefit 
\method. 
%
%
In addition, \method shows better stability in learning deep and wide models than the baseline methods. 
%
In terms of run-time performance, we also observed that \method is more efficient than the baselines, and thus 
has a great potential to be applied in systems with limited computing resources. 


%
%

\clearpage
\bibliographystyle{ACM-Reference-Format}
\bibliography{main}


\end{document}